\documentstyle[aps,preprint,tighten]{revtex}
\draft
\begin{document}
\title{Multifractal properties of critical eigenstates in two-dimensional
systems with symplectic symmetry}
\author{Ludwig Schweitzer}
\address{Physikalisch-Technische Bundesanstalt, Bundesallee
100, D-38116 Braunschweig, Germany}
\maketitle
\begin{abstract}
The multifractal properties of electronic eigenstates at the metal-insulator
transition of a two-dimensional disordered tight-binding model
with spin-orbit interaction are investigated numerically.
The correlation dimensions of the spectral measure $\widetilde{D}_{2}$
and of the fractal eigenstate $D_{2}$ are calculated and shown to be
related by $D_{2}=2\widetilde{D}_{2}$.
The exponent $\eta=0.35\pm 0.05$ describing the energy correlations of the
critical eigenstates is found to satisfy the relation $\eta=2-D_{2}$.
\end{abstract}

Electronic states in disordered systems are known to be either
localized or  extended. At $T=0$ the so called mobility edge
separates insulating  (localized) from current carrying (extended)
states.  Recently, the electronic properties directly at the critical point
have received  increasing attention\cite{CD88,Sea93,HS94,KLAA94,BSK94,HoSb94}.
Two-dimensional (2d) systems are very well suited for numerical investigations
of the electronic eigenstates at a critical point.
The only systems, however, that exhibit a complete Anderson transition
in 2d are models with symplectic symmetry.
This is in contrast to the orthogonal and the unitary case where all
states are localized (weak localization) and the quantum Hall (QHE) systems
where only localized states with a diverging localization length at some
singular energies can be observed.

In a recent paper, Chalker and coworkers \cite{CDEN93} reported on
numerical  investigations of eigenstate fluctuations and
correlations near the mobility  edge of a two-dimensional
tight-binding model with spin-orbit coupling. It was found that the
probability amplitude distribution  exhibits multifractal behavior
which can be characterized by a set of  generalized dimensions
$D_{q}$. The fractal spatial structure of the  wavefunctions also
shows up in the eigenfunction correlations between states  close in
energy. A similar behavior was found also in the quantum Hall model
\cite{CD88,HS92,PJ91,HKS92}, i.e. a two-dimensional disordered system of
non-interacting electrons where a strong magnetic field causes the
localization length to diverge at the centres of the Landau bands
with a universal exponent \cite{Huc94}.
For these critical states, characteristic quantities like the
generalized correlation dimension of the  wavefunction $D_{2}=1.62$
\cite{HS92,PJ91}, the corresponding generalized  dimension of the
spectral measure $\widetilde{D}_{2}=0.81$ \cite{HS94}, the  exponent
$\eta=0.38$ \cite{CD88,HS94} governing the energy eigenfunction
correlations, and $\alpha_{0}=2.3$ \cite{PJ91,HKS92} which for $-1\lesssim
q \lesssim 1$  determines the so called $f(\alpha)$-distribution are
known, partly with  sufficient precision. This is, however, not the
case for the spin-orbit system  mentioned above, mainly because of
the small system sizes considered so far ($M\le 18\,a) $\cite{CDEN93}.
The authors of Ref.~\cite{CDEN93} used a model proposed  by Evangelou and
Ziman\cite{EZ87} which represents a two-dimensional disordered electronic
system with spin-orbit interaction.

In this letter, a different lattice model with symplectic symmetry is
investigated for system sizes up to $150\,a\times 150\,a$, where $a$ is the
lattice constant. It is found that $D_{2}=1.66\pm0.05$,
$\widetilde{D}_{2}=0.83\pm 0.03$, $\eta=0.35\pm0.05$,
and $\alpha_{0}=2.19\pm 0.03$. Thus, within the numerical
uncertainties, the proposed relations
$\eta=2-D_{2}$ \cite{Cha90,PJ91}, $D_{2}=2\widetilde{D}_{2}$ \cite{HS94},
and $\Lambda^{\rm typ} _{0}\pi= (\alpha_{0}-2)^{-1}$ \cite{Jan94}
hold also for the two-dimensional system with
spin-orbit interaction.

The model used to calculate the eigenstates has been put forward by Ando
\cite{And88,And89} to simulate two-dimensional systems in $n$-channel
inversion layers on surfaces of III-VI semiconductors. The Hamiltonian
describing this situation is
\begin{equation}
{\cal H} = \sum_{m,\sigma } \varepsilon^{} _{m} c^{\dagger}_{m,\sigma }
c^{}_{m,\sigma '} + \sum_{mn, \sigma \sigma '}
V(m,\sigma ; n,\sigma ')\,c^{\dagger}_{m,\sigma } c^{}_{n,\sigma '},
\end{equation}
with disorder potentials $\varepsilon _{m}$, creation
$c^{\dagger}_{m,\sigma }$ and annihilation $c^{}_{m,\sigma }$
operators of a particle at site $m$ and spin state $\sigma $, respectively.
The transfer matrix elements $V(m,\sigma ; n,\sigma ')=
\sigma \sigma '\,V(n,-\sigma '; m,-\sigma )$ which are restricted to nearest
neighbours only depend on whether the transfer from site $m$ to one
of the nearest neighbours $n$ goes along the $x$- or the $y$-direction.
The strength of the spin-orbit interaction is determined by the parameter
$S=V_{2}/V$, with $V=(V^{2}_{1}+V^{2}_{2})^{1/2}$ taken to be the unit of
energy,
where $V_{1}$ and $V_{2}$ are the matrix elements for transitions with and
without spin-flip, respectively. $V(m,\sigma ; n\sigma ')$ is then given by
\begin{equation}
V_{x}=\left(
\begin{array}{ll}
V_{1} & V_{2} \\
-V_{2}& V_{1}
\end{array} \right), \
V_{y}=\left(
\begin{array}{ll}
V_{1} & {\rm i} V_{2} \\
{\rm i} V_{2} & V_{1}
\end{array} \right), \hspace{.5cm}
|\sigma =+1\rangle =\left(
\begin{array}{c}
1\\0
\end{array}\right), \
|\sigma =-1\rangle =\left(
\begin{array}{c}
0\\1
\end{array}\right).
\end{equation}
The localization properties of symplectic models have previously been
analyzed numerically \cite{EZ87,And88,And89,Mac85,Mac90,Fea91,Fas92}
from which a metal-insulator transition can be inferred. The most recent
calculations of Fastenrath \cite{Fas92,Fas92a} report a critical exponent
$\nu = 2.75$ for the localization length at the band centre, $E/V=0$,
together with a critical disorder $W_{c}=5.74\,V$ for a constant probability
distribution of the on-site disorder potentials $\{\varepsilon _{m}\}$
and a spin-orbit strength $S=0.5$. We also take these parameters
in what follows so that our new results supplement the already published
data. The eigenvalues and eigenstates were calculated numerically by means of
a Lanczos-algorithm for systems of size up to $L=150\,a$ with periodic
boundary conditions applied in both directions.

The structure of the eigenstates is analyzed in terms of the
$f(\alpha)$-distribution\cite{Hea86,CJ89} which completely characterizes
the spatial scaling behavior of the $q$-moments of the wavefunction.
In Fig.~1 the $f(\alpha)$-distribution for a particular eigenstate from the
critical region near the center of the tight binding band ($E=0\,V$) and a
disorder strength $W=5.74\,V$ is shown for several $q$-values. For comparison
the corresponding parabolic approximation\cite{CW87,Jan94} is also shown.
The average over 190 eigenstates taken from the energy interval $[-0.15,0.0]$
gives a value of $\alpha_{0} = 2.19\pm 0.03$. This number represents the most
probable value of the scaling exponents
$\alpha_{q}= \mbox{d}/\mbox{d}q\,(q-1) D_{q}$.

Using the box-probability method the correlation dimension $D_{2}$ of the
multifractal eigenstates is obtained from the scaling of the second moment
(q=2) of the averaged box-probability $P(q,\lambda)=
\sum_{i}(\sum_{r\in\Omega_{i}(\lambda)}|\psi(r)|^2)^q \sim
\lambda ^{(q-1)D_{q}}$,
where $\Omega_{i} (\lambda )$ is the size $l\times l$ of the $i$-th box.
A power law relation is observed for $l=\lambda L$ in the range $2a \le l
\le L/2$. The average over all eigenstates from the energy interval gives
a correlation dimension $\langle D_2 \rangle_{E} = 1.66 \pm 0.05$ which is
close to the value of 1.63 reported in \cite{CDEN93} for smaller system sizes.

The correlation dimension of the spectral measure, $\widetilde{D}_{2}$,
which is related to the temporal decay\cite{KPG92,HS94} of the maximum
of a wavepacket (probability of return) built from critical eigenstates
was calculated for the same energy interval. This exponent is obtained from
the scaling relation of the local density of states
\begin{equation}
\gamma(q,\varepsilon )=\lim_{\varepsilon \to 0}1/L^{2}\sum_{{\bf r}} \sum_{i}
\Big(\sum_{E \in \Omega _{i}(\varepsilon) } |a_{E}|^{2}\Big)^{q}
\sim \varepsilon ^{(q-1)\widetilde{D}_{q}},
\end{equation}
with $a_{E}=\psi _{E}({\bf r})/(\sum_{E'}|\psi _{E'}({\bf r})|^{2})^{1/2}$.
The scaling behavior is shown in Fig.~2 for $q=2$.
Here, an exponent $\widetilde{D}_{2}=0.83\pm 0.03$ is found
which satisfies the relation $D_2=2\widetilde{D}_{2}$ proposed previously for
the 2d QHE system\cite{HS94}. These results indicate that the diffusion at the
mobility edge will be non-Gaussian in the long-time limit. Due to the
multifractal nature of the spatial amplitude fluctuations of the critical
wavefunctions and the behavior of the fractal spectral measure, the diffusion
coefficient will not be a constant. A non-trivial frequency and wavevector
dependence governed by an exponent $\eta$ has been observed in the
QHE model \cite{CD88,HS94} and a similar behavior was also seen for the
symplectic case \cite{CDEN93}.

In Fig.~3 the correlations of the eigenstates close in energy
\begin{equation}
Z(E,E')=
\sum_{\bf r} |\psi_E({\bf r})|^2 |\psi_{E'}({\bf r})|^2 \sim |E-E'|^{-\eta/d}
\end{equation}
averaged over small energy intervals are shown for the critical eigenstates.
A power law relation is observed with $\eta=0.35\pm 0.05$.
{}From the above results it is seen that the relation $\eta=d-D_2$
\cite{Cha90,PJ91} also holds in the 2d symplectic case.

We have to mention that our value for $\eta $ is compatible with the result
for a different symplectic model  obtained from the calculation
of the two-particle spectral function at the mobility edge \cite{CDEN93}.
It is also in accordance with an earlier estimate by Evangelou \cite{Eva90},
although the corresponding scaling exponent of the localization length,
$\nu=1.6$, differs considerably from $\nu =2.75$ obtained by Fastenrath
\cite{Fas92,Fas92a} for the Ando Model. However, to determine $\eta$ from the
calculated finite size scaling variable at the critical point,
$\Lambda _{c}$, Evangelou  used the relation $\Lambda _{c}=1/(\pi \eta/2)$
which is only within the parabolic approximation ($D_{q} = 2-q(\alpha_{0}-2)$
\cite{CW87,Jan94}, valid for $|q|\lesssim 1$)
equivalent to a relation proposed recently by Jan{\ss}en \cite{Jan94},
$\Lambda^{\rm typ} _{c}=1/(\pi (\alpha_{0}-d))$,
relating the typical finite size scaling variable at the critical point,
$\Lambda^{\rm typ}_{c} $, to the multifractal behavior of the eigenstates.

We note that our results for $\alpha_{0}$, $D_{2}$,
$\widetilde{D}_{2}$, and $\eta$ are very close to those obtained for the 2d
QHE model\cite{HKS92,HS94}.
Although we could not observe any size dependence, the presently achieved
system sizes do, however, not allow to exclude the possibility that the
values actually coincide in the thermodynamic limit, $L \to \infty$.
A similar correspondence has recently been asserted for the energy level
statistics \cite{HoSb94} of the critical 3d Anderson model with and without a
magnetic field. In addition, the critical exponent of the localization length
was reported to be identical in both cases\cite{OKO93,HK94}.
If these observations for 3d models were correct and if also the values
obtained from the multifractal analysis in 2d for the quantum Hall systems were
indeed the same as those reported above for the symplectic disordered systems,
then the behavior at the critical point would primarily be determined by the
euclidean dimension and not by the symmetry class of the Hamiltonian.

In conclusion, the multifractal properties of the electronic eigenstates
at the metal-insulator transition of two-dimensional disordered systems with
symplectic symmetry have been investigated. The $f(\alpha)$-distribution,
the correlation dimensions of the spectral measure and the critical
eigenstates, and the energy correlations of the wavefunctions were calculated.
The obtained values for $\alpha_0=2.19\pm 0.03$,
$\widetilde{D}_{2}=0.83\pm 0.03$, $D_2=1.66\pm 0.05$, and $\eta=0.35\pm 0.05$
appear to be independent of system size
and satisfy general relations that have been proposed previously for the
critical states at the metal-insulator transition in other disordered systems.

\acknowledgements
I gratefully acknowledge helpful discussions with Bodo Huckestein and Martin
Jan{\ss}en.

\bibliographystyle{prsty}

\begin{thebibliography}{99}

\bibitem{CD88}
J.~T. Chalker and G.~J. Daniell, Phys.\ Rev.\ Lett. {\bf 61},  593  (1988).

\bibitem{Sea93}
B.~I. Shklovskii {\it et~al.}, Phys.\ Rev.\ B {\bf 47},  11487  (1993).

\bibitem{HS94}
B. Huckestein and L. Schweitzer, Phys.\ Rev.\ Lett. {\bf 72},  713  (1994).

\bibitem{KLAA94}
V.~E. Kravtsov, I.~V. Lerner, B.~L. Altshuler, and A.~G. Aronov, Phys.\ Rev.\
  Lett. {\bf 72},  888  (1994).

\bibitem{BSK94}
T. Brandes, L. Schweitzer, and B. Kramer, Phys.\ Rev.\ Lett. {\bf 72},  3582
  (1994).

\bibitem{HoSb94}
E. Hochstetter and M. Schreiber, Phys.\ Rev.\ Lett. {\bf 73},  3137  (1994).

\bibitem{CDEN93}
J.~T. Chalker, G.~J. Daniell, S.~N. Evangelou, and I.~H. Nahm, J. Phys.:\
  Condens.\ Matter {\bf 5},  485  (1993).

\bibitem{HS92}
B. Huckestein and L. Schweitzer,  in {\em High Magnetic Fields in Semiconductor
  Physics III: Proceedings of the International Conference, W\"u{}rzburg 1990},
  edited by G. Landwehr (Springer Series in Solid-State Sciences 101, Springer
  Verlag, Berlin, 1992), p.\ 84.

\bibitem{PJ91}
W. Pook and M. Jan{\ss}en, Z. Phys.\ B {\bf 82},  295  (1991).

\bibitem{HKS92}
B. Huckestein, B. Kramer, and L. Schweitzer, Surf.\ Science {\bf 263},  125
  (1992).

\bibitem{Huc94}
B. Huckestein, Phys.\ Rev.\ Lett. {\bf 72},  1080  (1994).

\bibitem{EZ87}
S.~N. Evangelou and T.~A.~L. Ziman, J. Phys.:\ Condens.\ Matter {\bf 20},  L235
   (1987).

\bibitem{Cha90}
J.~T. Chalker, Physica A {\bf 167},  253  (1990).

\bibitem{Jan94}
M. Jan{\ss}en, International Journal of Modern Physics B {\bf 8},  943  (1994).

\bibitem{And88}
T. Ando, Surf.\ Science {\bf 196},  120  (1988).

\bibitem{And89}
T. Ando, Phys.\ Rev.\ B {\bf 40},  5325  (1989).

\bibitem{Mac85}
A. MacKinnon,  in {\em Localization, Interactions and Transport Phenomena},
  edited by B. Kramer, G. Bergmann, and Y. Bruynseraede (Springer Verlag,
  Berlin, 1985), p.\ 90.

\bibitem{Mac90}
A. MacKinnon,  in {\em Localization and Confinement of Electrons in
  Semiconductors}, Vol.~97 of {\em Springer Series in Solid State Sciences},
  edited by F. Kuchar (Springer Verlag, Berlin, 1990), pp.\ 111--116.

\bibitem{Fea91}
U. Fastenrath {\it et~al.}, Physica A {\bf 172},  302  (1991).

\bibitem{Fas92}
U. Fastenrath, Helvetica Physica Acta {\bf 65},  425  (1992).

\bibitem{Fas92a}
U. Fastenrath, Physica A {\bf 189},  27  (1992).

\bibitem{Hea86}
T.~C. Halsey {\it et~al.}, Phys.\ Rev.\ A {\bf 33},  1141  (1986).

\bibitem{CJ89}
A. Chhabra and R.~V. Jensen, Phys.\ Rev.\ Lett. {\bf 62},  1327  (1989).

\bibitem{CW87}
M.~E. Cates and T.~A. Witten, Phys.\ Rev.\ A {\bf 35},  1809  (1987).

\bibitem{KPG92}
R. Ketzmerick, G. Petschel, and T. Geisel, Phys.\ Rev.\ Lett. {\bf 69},  695
  (1992).

\bibitem{Eva90}
S.~N. Evangelou, J. Phys. A: Math. Gen. {\bf 23},  L317  (1990).

\bibitem{OKO93}
T. Ohtsuki, B. Kramer, and Y. Ono, J. Phys. Soc. Jpn. {\bf 62},  223  (1993).

\bibitem{HK94}
M. Hennecke, B. Kramer, and T. Ohtsuki, Europhys. Lett {\bf 27},  389  (1994).

\end{thebibliography}

\begin{figure}[htb]
\caption[]{The $f(\alpha (q))$-distribution function of critical eigenstates
calculated for a system of size $L/a=150$ and $q=\pm
3$, $\pm 2$, $\pm1.5$, $\pm 1$, $\pm 0.8$, $\pm0.5$, $\pm 0.3$, 0.
The full curve is a fit using the parabolic approximation with
$\alpha _{0}=2.19$.}
\end{figure}

\begin{figure}[htb]
\caption[]{The scaling of the spectral measure at the mobility edge for a
system of size $L/a=100$ from which a correlation exponent
$\widetilde{D}_{2}=0.83 \pm 0.03$ is obtained.}
\end{figure}

\begin{figure}[htb]
\caption[]{The energy correlation $Z(E,E')$ of the critical eigenstates
as a function of energy separation showing a power law relation
$\sim |E-E'|^{-\eta/2}$ with an exponent $\eta=0.35\pm 0.05$.}
\end{figure}
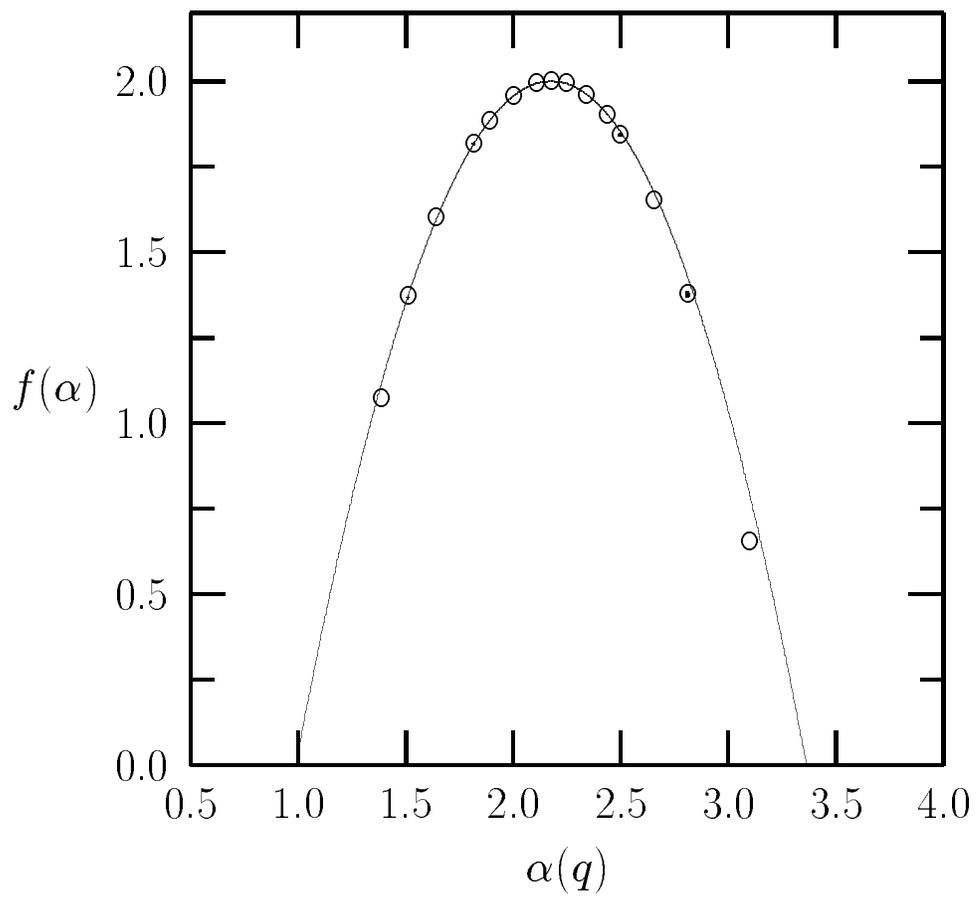
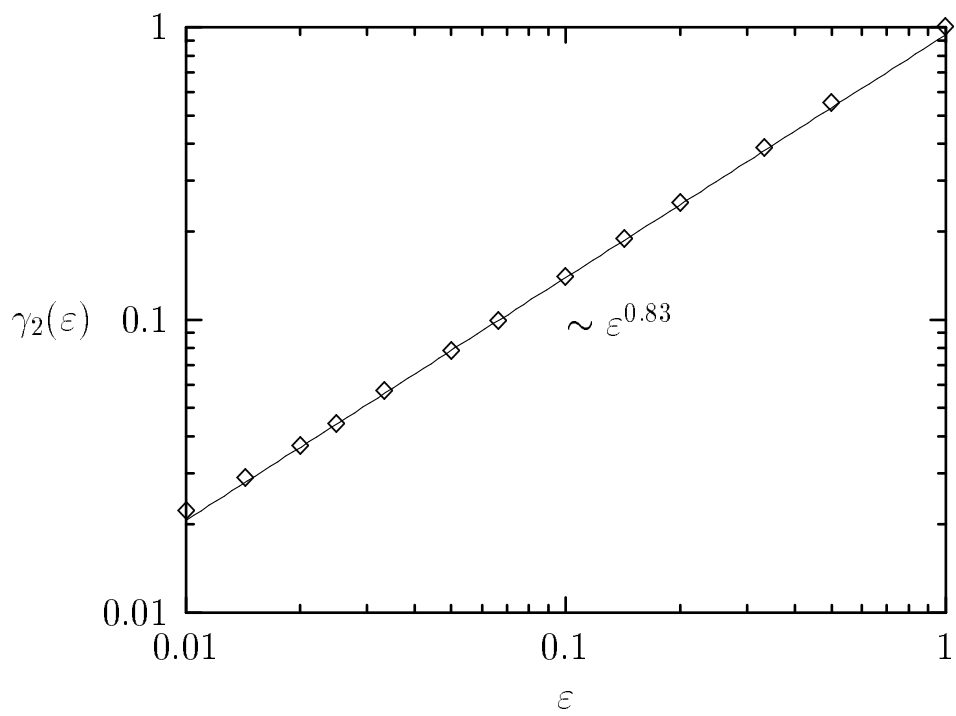
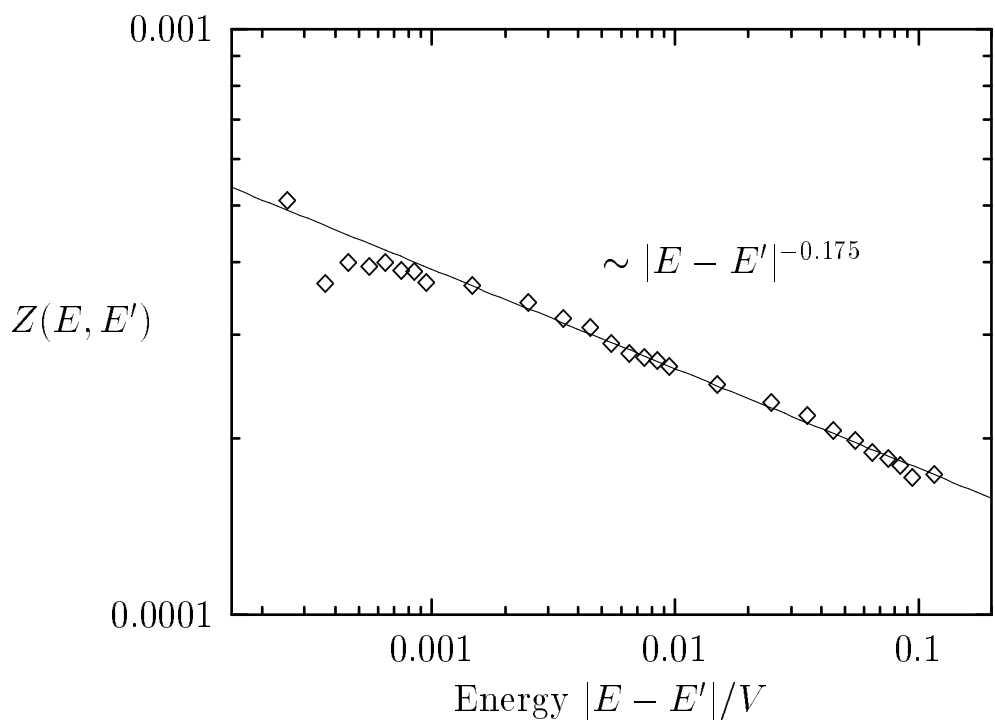
\end{document}